\documentclass[twocolumn,prl,preprintnumbers,amsmath,amssymb,groupedaddresst]{revtex4}
\usepackage{graphicx}
\usepackage{color}
\begin{document}

\preprint{RIKEN-MP-52}
\preprint{RIKEN-QHP-31}
\preprint{YGHP-12-49}

\title{Ferromagnetic neutron stars: \\
axial anomaly, dense neutron matter, and pionic wall}

\author{Minoru  {\sc Eto}$^{*}$}
\author{Koji {\sc Hashimoto}$^\dagger$}
\author{Tetsuo  {\sc Hatsuda}$^+$}
\affiliation{$*$ {\it Department of Physics, Yamagata University, Yamagata 990-8560, Japan}}
\affiliation{
$\dagger${\it Mathematical Physics Lab., RIKEN Nishina Center, Saitama 351-0198, Japan }}
\affiliation{
$+${\it  Theoretical Research Division, RIKEN Nishina Center, Saitama 351-0198, Japan }}

\begin{abstract}
We show that a  chiral nonlinear sigma model coupled to degenerate neutrons exhibits
a ferromagnetic phase at high density. The magnetization is due to the 
axial anomaly acting on the 
 parallel layers of neutral pion domain walls spontaneously formed at high density.
The  emergent magnetic field would reach the QCD scale $\sim 10^{19}$ [G],
 which suggests that the quantum anomaly can be  
a microscopic origin of the magnetars (highly magnetized neutron stars).
\end{abstract}

\maketitle


\vspace{2mm}
{\noindent \bf Magnetars and high density neutron matter.} ---
The phase diagram of QCD is a mystery to be uncovered. Even though
the problem is theoretically well-posed with QCD Lagrangian, so far the difficulty 
remains yet to reveal a specific region of the phase diagram, the region 
at low temperature and   high baryon number density $\rho$  \cite{Fukushima:2010bq}.

 One of the promising systems where such a high density  
 region is realized in nature is  the deep interior of compact stars, such as the 
 neutron stars.  Observations of various properties of these stars
  should give us crucial constraint on  high density matter.
 In particular, the magnetars, which are considered to be neutron stars with 
  very strong magnetic field $\sim 10^{15}$[G] at their surface, are of partiular interests
  \cite{EnotoProc,Mereghetti:2008}. 
 The mechanism for generating such a strong magnetic field is so far not established: 
 Among various proposals including the 
  the  dynamo formation model and the fossil-field model  \cite{Mereghetti:2008},
 nuclear ferromagnetism associated with the 
 solidification of the neutron star core, which was originally proposed right after
  the discovery of the pulsar  \cite{Brownell:1969},  is an  interesting possibility
   to generate large intrinsic magnetic field.
 However,  
   modern  quantum many-body calculations on the neutron matter and asymmetric nuclear matter
  with realistic nuclear force
  have shown that these systems stay in the liquid phase at high density 
   without having  spontaneous ferromagnetic transition  \cite{Bordbar:2008}.
  Another interesting possibility is  the ferromagnetism of the 
   quark liquid in the central core of neutron star \cite{Tatsumi:1999ab};  
  spin-polalized quark phase, similar to that in the low density electron gas,
   may be realized in a certain window of baryon density
       due to the   Fock term of  the gluon exchange between quarks. 

In this letter, we propose a novel mechanism which leads to a {\em spontaneous}
 magnetization
of the neutron matter, based on the non-linear chiral Lagrangian of pions
coupled to degenerate neutrons. Two basic ingredients are 
(i) the neutron spin-density induced on a pion domain wall
in dense matter \cite{Hatsuda:1986nq} and (ii)
the baryon number induced on the pion domain wall by an external magnetic field through axial
 anomaly  \cite{Son:2007ny}.
We show that layers of $\pi^0$ domain walls are spontaneously generated 
 by a small seed of an external magnetic field.  Then, 
  intrinsic magnetic field is induced from the layers which have 
  net magnetization. 
 If this mechanism takes place inside the core of the neutron stars above certain
  threshold density, they acquire large magnetic field and become  magnetars.


\vspace{2mm}
\noindent
{\bf Chiral Lagrangian and pion domain walls.} ---
We use the chiral Lagrangian
for low energy pions and nucleons with the Weinberg parametrization \cite{Weinberg:1968de}:
\begin{eqnarray}
{\cal L} &=&
 \bar{N} 
 \bigl[  i  \gamma^\mu 
 \left(  \partial_\mu 
   + i \mbox{\boldmath$\tau$} \cdot  \mbox{\boldmath $V$}_{\mu}
   + i \gamma_5 \mbox{\boldmath$\tau$} \cdot  \mbox{\boldmath $A$}_{\mu} 
 \right)  
  - m_{_{\rm N}}  
 \bigr] N
\nonumber \\
& & 
 +  \frac{1}{2}| D_\mu \mbox{\boldmath $\phi$}|^2
 - \frac{1}{2} (m_{\pi}^2 + \sigma_{\pi N} \bar{N} N)
  \frac{\mbox{\boldmath $\phi$}^2}{1 + {\mbox{\boldmath $\phi$}^2}/{4f_\pi^2}}.\;\;
\end{eqnarray}
Here 
$N$ is the nucleon field (isospin $SU(2)$ doublet), $\mbox{\boldmath $\phi$}$ is
the pion (triplet),
$f_\pi$ is the pion decay constant, $g_A$ is the axial charge of the nucleon,
and $m_\pi$ is the pion mass. Also,
$\mbox{\boldmath $V$}_{\mu} \equiv \frac{1}{4f_{\pi}^2} (1+
 \mbox{\boldmath $\phi$}^2/4f_\pi^2)^{-1} (\mbox{\boldmath $\phi$} 
 \times \partial_{\mu} \mbox{\boldmath $\phi$})$,
 $\mbox{\boldmath $A$}_{\mu} \equiv 
  \frac{g_A}{2 f_{\pi}}      D_\mu \mbox{\boldmath$\phi$}$  
with  $D_\mu \mbox{\boldmath $\phi$} \equiv
(1 + \mbox{\boldmath $\phi$}^2/4f_\pi^2)^{-1} \partial_\mu \mbox{\boldmath $\phi$}$, 
respectively.

\begin{figure}[t]
\begin{center}
\includegraphics[width=0.35\textwidth]{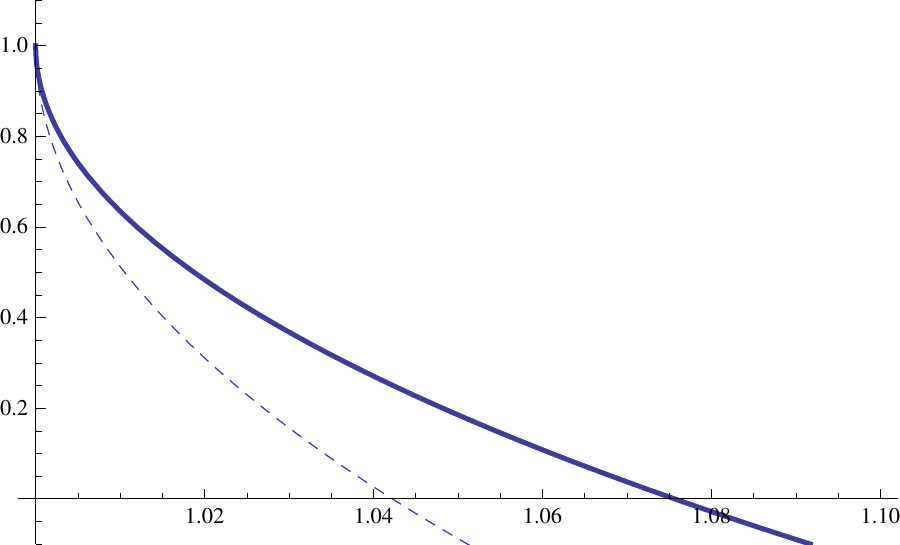}
\put(0,15){$\mu_{\rm n}/m_{\rm N}$}
\put(-110,45){$\beta$} 
\put(-140,20){$\tilde{\beta}$} 
\put(-169,0){1.00}
\end{center}
\vspace{-5mm}
\caption{Plots of $\beta$ (rigid lines) and $\tilde{\beta}$ (dashed lines) as
a function of $\mu_{\rm n}$.
$\beta$ decreases monotonically as the density grows. Here we used 
$m_{_{\rm N}}=940$ [MeV], $f_\pi = 93$[MeV] and $g_A=1$ 
(with which $\beta$ vanishes at $\mu_{\rm n} \sim 1.08 m_{\rm N}$
corresponding to $\rho_n \sim 0.23$ [fm$^{-3}$]). In general in dense matter
$m_{_{\rm N}}$ and $f_\pi$ may vary, so the plot should be understood only qualitatively.
$\alpha$ and $\tilde{\alpha}$ do not vary much in density. 
}
\label{fighat}
\end{figure}
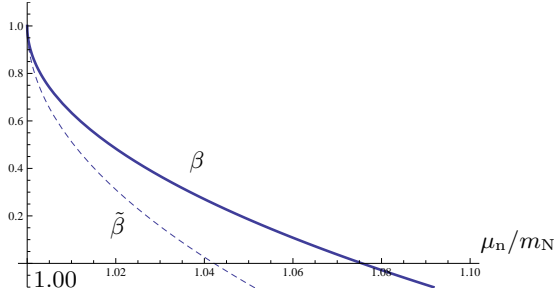

In \cite{Hatsuda:1986nq}, neutrons are integrated with neutron chemical potential,
and $\pi^0$ domain wall solutions were studied. 
Here we generalize
it to include the charged pion and obtain the in-medium 
chiral Lagrangian up to $O(p^2)$;
\begin{eqnarray}
&&{\cal L}_{\rm eff} =
\frac{\alpha}{2} |D_0 \phi_3|^2 - \frac{\beta}{2} |D_i \phi_3|^2
+  \frac{\tilde{\alpha}}{2} |D_0 \phi_+|^2 - \frac{\tilde{\beta}}{2} | D_i \phi_+|^2 
\nonumber \\
&&
 \quad -\gamma_0 \frac{m_{\pi}^2}{2}  
  \mbox{\boldmath $\phi$}^2/\left( 1 + {\mbox{\boldmath $\phi$}^2}/{4f_\pi^2} \right) 
  \, \nonumber 
  \\
  &&
\quad  + \gamma_1 \; (\phi_+ (D_0 \phi_+)^*\!\!-\!(\phi_+)^* D_0 \phi_+)
\label{corac}
 \\ &&
\quad   + \gamma_2 \; (\phi_+ (D_0 \phi_+)^*\!\!-\!(\phi_+)^* D_0 \phi_+)^2
    \nonumber \\ &&
 \quad + \gamma_3 \; |\phi_+ D_0 \phi_3 \!-\! \phi_3 D_0 \phi_+|^2
 + \gamma_4 \; |\phi_+ D_i \phi_3 \!-\! \phi_3 D_i \phi_+|^2
\nonumber
\end{eqnarray}
with $i=1,2,3$.
We have defined the charged pion $\phi_+ \equiv \phi_1 + i \phi_2$, and $D_\mu \phi_+ \equiv
(\partial_\mu \phi_+ + i \delta_{\mu 0} \mu_{\rm I}\phi_+ )/(1 + |\mbox{\boldmath $\phi$}|^2/4f_\pi^2)$,
where $\mu_{\rm I}$ is the isospin chemical potential defined as the
difference $\mu_{\rm I} \equiv \mu_{\rm p}-\mu_{\rm n}$.
The correction due to the background neutrons is in the coefficients 
$\alpha, \beta, \tilde{\alpha}, \tilde{\beta}$ (and $\gamma_{1,2,3,4}$), 
which are equal to the unity (zero) in the absence of 
the nucleons. 
They are given by one-loop calculations (see \cite{Hatsuda:1986nq} for $\alpha$ and $\beta$)
\begin{eqnarray}
\beta &\equiv&  1-\frac{g_A^2m_{_{\rm N}}^2}{4\pi^2f_\pi^2}
\log(x + \sqrt{x^2-1}) ,
\nonumber
\\ 
\alpha &\equiv &  1+\frac{g_A^2m_{_{\rm N}}^2}{4\pi^2f_\pi^2}
 (x\sqrt{x^2-1} - \log (x + \sqrt{x^2-1})),
\nonumber
\end{eqnarray}
\begin{eqnarray}
\tilde{\beta} &\!\equiv\!& 
 1\!-\!\frac{g_A^2 m_{_{\rm N}}^2}{2\pi^2f_\pi^2}\!\!
\int_1^x \!\!\!\!\! ds \frac{(s^2\!-\!1)^{1/2} (2s^2\!+\! 4 \!+\! 3(1\!-\!x) s)}{3(1-x+2s)(x\!-\!1)} ,
\nonumber
\\ 
\tilde{\alpha} &\equiv & 
 1\!+\!\frac{g_A^2 m_{_{\rm N}}^2}{2\pi^2f_\pi^2}
\int_1^x \!\!\!\!\! ds \frac{(s^2\!-\!1)^{1/2} (2s^2\!-\! 2 \!+\! (1\!-\!x) s)}{(1-x+2s)(x\!-\!1)} ,
\nonumber
\\ 
\gamma_0 &\equiv & 
 1 \!+\! \frac{ \sigma_{\pi {\rm N} } }{m_{\pi}^2}
  \frac{m_{_{\rm N}}^3}{2\pi^2} (x\sqrt{x^2\!-\!1} \!-\! \log (x\! +\! \sqrt{x^2\!-\!1})),
\nonumber
\\
\gamma_1 &\equiv &
\frac{im_{_{\rm N}}^2}{24\pi^2 f_\pi^2}  (x^2\!-\!1)^{3/2} , \;\;
\gamma_2 \equiv 
\frac{ -m_{_{\rm N}}^2}{128\pi^2 f_\pi^4} x\sqrt{x^2\!-\!1},
\nonumber \\
\gamma_3 &\equiv &
\frac{m_{_{\rm N}}^2}{16\pi^2f_\pi^4}
\int_1^x \!\!\! ds \frac{(s^2\!-\!1)^{1/2} (2s^2\! +\! (1\!-\!x) s)}{(1-x+2s)(x\!-\!1)} ,
\nonumber
\nonumber \\
\gamma_4 &\equiv &
\frac{m_{_{\rm N}}^2}{16\pi^2f_\pi^4}
\int_1^x \!\!\! ds \frac{(s^2\!-\!1)^{1/2} (2s^2\! -\!2\!+\!3 (1\!-\!x) s)}{3(1-x+2s)(x\!-\!1)} .
\nonumber
\end{eqnarray}
Here $x\equiv \mu_n/m_{_{\rm N}}$, and 
we consider the pure neutron matter for simplicity, so that 
$\mu_{\rm p}=m_{_{\rm N}}$. 
The neutron density is related to the Fermi momentum as $\rho_n = k_{\rm F}^3/(3\pi^2)$ 
with $k_{\rm F}\equiv \sqrt{\mu_n^2-m_{_{\rm N}}^2}$. 
The $\sigma_{\pi{\rm N}}$ correction is taken care of at its leading order.
An important feature of the corrections is that $\beta$ is a monotonically decreasing
function of the neutron density, and vanishes 
at a certain density (Fig.~\ref{fighat}).

A classical solution of Eq.\eqref{corac} is a domain wall of the in-medium
 neutral pion \cite{Hatsuda:1986nq},
\begin{eqnarray}
\phi_3 = \frac{2 f_\pi}{\sinh [ m_\pi x^3/\sqrt{\beta/\gamma_0}]}, \quad \phi_1=\phi_2=0.
\label{sol}
\end{eqnarray}
Note that this $\pi^0$ domain wall interpolates the vacua $\theta=0$ and $\theta = 2\pi$,
where 
$\tan \frac{\theta}{2} \equiv |\mbox{\boldmath $\phi$}| /2 f_\pi$.
 Interestingly, the domain wall can reduce its weight in the neutron matter:
$\gamma_0$ stays positive at all $x$, while $\beta$ approaches zero as shown in Fig.1, so that 
the tension ${\cal E}/S$ of the domain wall is significantly reduced for $\beta \rightarrow 0$,
\begin{eqnarray}
{\cal E}/S = 8 \sqrt{\beta \gamma_0}f_\pi^2 m_\pi,
\label{tension}
\end{eqnarray}
where $S$ is the domain wall area.
 We will show that the parallel layers of the domain walls would populate at high
 baryon density due to the reduction of the tension and the accumulation of 
  baryon number on the domain wall. 

\begin{figure}[t]
\begin{center}
\includegraphics[width=0.3\textwidth]{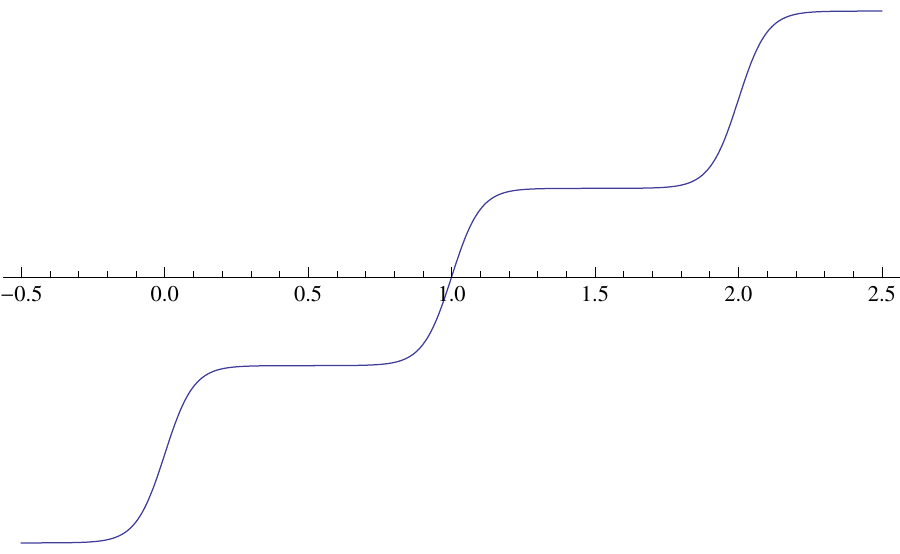}
\put(-10,55){$x^3$}
\put(-40,70){$\theta$}
\end{center}
\vspace{-7mm}
\caption{The multi domain wall solution written by $\theta(x^3)$.
Each wall interpolates adjacent vacua, $\theta = 0,2\pi,4\pi,\cdots$.
For the exact solution, see for example \cite{Eto:2004zc}.
}
\label{figphi3}
\end{figure}


\vspace{2mm}
{\noindent \bf Emergent magnetic field on pionic walls from axial anomaly.} ---
Let us first discuss how the spontaneous magnetization occurs in high density neutron 
matter. There are three steps for this to happen:
 (i) The in-medium domain wall becomes light,  and at the same time it acquires
  finite baryon density due to axial anomaly under an external magnetic field.
 Then the system with a domain wall 
  becomes energetically favorable than the uniform neutron matter above a certain density. 
  (ii)  The domain wall is magnetized due to the spin alignment of surrounding
  neutrons, so that it creates spontaneous  magnetization which enlarges the original
   magnetic field.
 (iii)  The enhanced magnetic field creates more domain walls. 
 The cycle (i)$\rightarrow$(ii)$\rightarrow$(iii)$\rightarrow$(i)
   is repeated and leads to a stable configuration with many parallel layers
 of thin domain walls with a high magnetic field. All the spins are aligned, so the system is ferromagnetic.

We now explain the mechanism of inducing the baryon charge following Ref.\cite{Son:2007ny}.
In a background constant magnetic field, the QCD axial anomaly term in the 
 chiral Lagrangian reads 
\begin{eqnarray}
{\cal L}_{\rm WZW} = \frac{i e}{16\pi^2} A_0^{(B)} B_3 {\rm tr}\left[
\tau_3 (U \partial_3 U^\dagger + \partial_3 U^\dagger U)
\right],
\label{wzw}
\end{eqnarray}
where $A_0^{(B)}$ is the temporal component of 
a gauge potential for the baryon number symmetry,
$B_3$ is the background magnetic field along $x^3$, and 
$U \equiv \cos\theta + i \mbox{\boldmath $\tau$}\cdot 
\hat{\mbox{\boldmath $\phi$}} \sin\theta$
 with 
$\hat{\mbox{\boldmath $\phi$}}
=\mbox{\boldmath $\phi$} /|\mbox{\boldmath $\phi$}|$. 
Since the pion-dependent part can be evaluated (for $\phi_1=\phi_2=0$) as
\begin{eqnarray}
{\rm tr}\left[
\tau_3 (U \partial_3 U^\dagger + \partial_3 U^\dagger U)\right]
= -4i \frac{D_3 \phi_3}{f_\pi} = -4i \partial_3 \theta,
\end{eqnarray}
we immediately see that the domain wall, which interpolates $\theta=0$ and $\theta=2\pi$, 
can obtain a baryon charge per a unit area \cite{Son:2007ny},
\begin{eqnarray}
N_B/S = e B_3/2\pi. 
\label{baryond}
\end{eqnarray}
If the domain wall is not parallel to the magnetic field, the induced baryon charge is
reduced to $B_i \hat{n}_i$ where $\hat{n}$ is the unit vector perpendicular to the 
domain wall. Note that the formula Eq.\eqref{baryond} is valid even for  $\beta\neq 1$.

Combining  Eq.\eqref{baryond} with Eq.\eqref{tension}, 
the domain wall energy per a unit baryon charge is
\begin{eqnarray}
{\cal E}/N_B = \frac{16\pi f_\pi^2 m_\pi}{e  B_3} \sqrt{\beta \gamma_0}.
\label{EoverN}
\end{eqnarray}
The system with domain wall is more favorable than the uniform neutron matter 
when the domain wall energy per baryon becomes smaller than the 
 neutron chemical potential \cite{Son:2007ny}, i.e.
  $B_3 > \sqrt{\beta \gamma_0} \times (16 \pi f_{\pi}^2  m_{\pi}/\mu_{\rm n}) 
  \sim \sqrt{\beta \gamma_0} \times  10^{19} $[G].  
Here, the factor $\beta$, which was not taken into account in \cite{Son:2007ny},  is important.
 Since $\beta(\rho)$ is a monotonically decreasing function of $\rho_{\rm n}$,
an adiabatic increase of the density $\rho_{\rm n}$ inevitably hits the critical value of $\beta$ 
at which the domain wall is created. 
 
 Once the pionic wall is formed,
 neutron spins on the wall align
in the direction perpendicular to the domain wall \cite{Hatsuda:1986nq}.
The spin density of the neutrons is a spatial part of the axial current 
$j^{(A)}_i =\langle \bar\psi_n \gamma_i \gamma_5 \psi_n\rangle$.
It was evaluated in \cite{Hatsuda:1986nq}, in the same approximation, as
\begin{eqnarray}
s_3/S = 2\pi (\beta-1) f_\pi^2, 
\label{walltension}
\end{eqnarray}
which is the third component of the neutron 
spin density per a unit area of the domain wall.

We note here that neutrons have a magnetic moment $\mu =  g e s_3/2m_{\rm N}$
where $g\sim -3.8$ is the neutron $g$-factor. So the total magnetic moment (which is
the magnetization $M$) per a unit volume at $\beta \sim 0$ is
\begin{eqnarray}
M = \frac{\pi |g| e f_\pi^2}{m_{_{\rm N}}} \frac{1}{d}
\label{mag}
\end{eqnarray}
where $d$ is the separation between adjacent domain walls.
Therefore {\it the domain wall phase is ferromagnetic}.

The magnetization $M$ is larger for smaller separation $d$ among the domain walls.
This $d$ is intimately related to the induced baryon density due to the domain wall,
and in fact this is a driving force for developing strong magnetic field. From Eq.\eqref{baryond},
we know that the averaged baryon number density induced by the domain walls is
\begin{eqnarray}
\rho_{\rm dw} = \frac{e B_3}{2\pi} \frac{1}{d}.
\label{indb}
\end{eqnarray}
The total baryon number density $\rho$ is given as a sum of $\rho_{\rm dw}$ and
the remaining neutron density $\rho_n$. (This $\rho_n$ cannot vanish, since the $\beta$ correction needs background neutrons.) Combining Eq.\eqref{indb} with Eq.\eqref{mag}, we obtain
\begin{eqnarray}
M = \frac{2\pi^2 |g|  f_\pi^2 \rho_{\rm dw} }{m_{_{\rm N}}B_3}.
\label{mag2}
\end{eqnarray}
Once the background $B_3$ becomes larger, the domain wall energy cost
\eqref{EoverN} becomes smaller. So the domain wall is created easier, and 
the domain wall separation $d$ becomes smaller. Then from Eq.\eqref{mag}, the magnetization
$M$ increases and helps the $B_3$ to increase.
So, this system has a self-enhancement mechanism of the magnetic field.
The equilibrium can be reached at $B_3=M$ which is our critical induced
magnetic field,
\begin{eqnarray}
B_3 = \sqrt{2\pi^2 |g| f_\pi^2 \rho_{\rm dw} /m_N} \sim 3 \times 10^{19} \; \mbox{[G]}.
\label{b3cri}
\end{eqnarray}
supposing a typical value for $\rho\sim\rho_{\rm dw}\sim\rho_n$.
The value of the magnetic field 
is quite large ($\sqrt{B_3} \sim 10^2$ [MeV]) and close to the
QCD deconfinement scale around which our approximation breaks down\footnote{We can estimate
from these equations that the equilibrium magnetic field corresponds to $d\sim 0.1$[fm], at which
our low-energy approximation is not applicable.}.
The magnetization is expected to stop increasing somewhere before
reaching this value.

In addition to the neutron spin alignment, the WZW term \eqref{wzw} itself may provide
a magnetization \cite{Son:2004tq} of the same sign. Details will be in our forthcoming paper \cite{ours}.

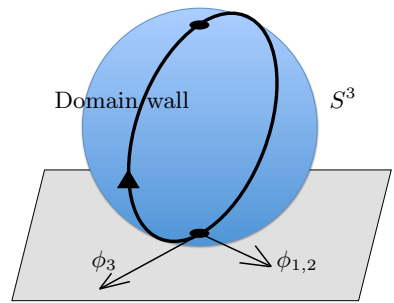
\begin{figure}[t]
\begin{center}
\vspace*{-5mm}
\includegraphics[width=0.4\textwidth]{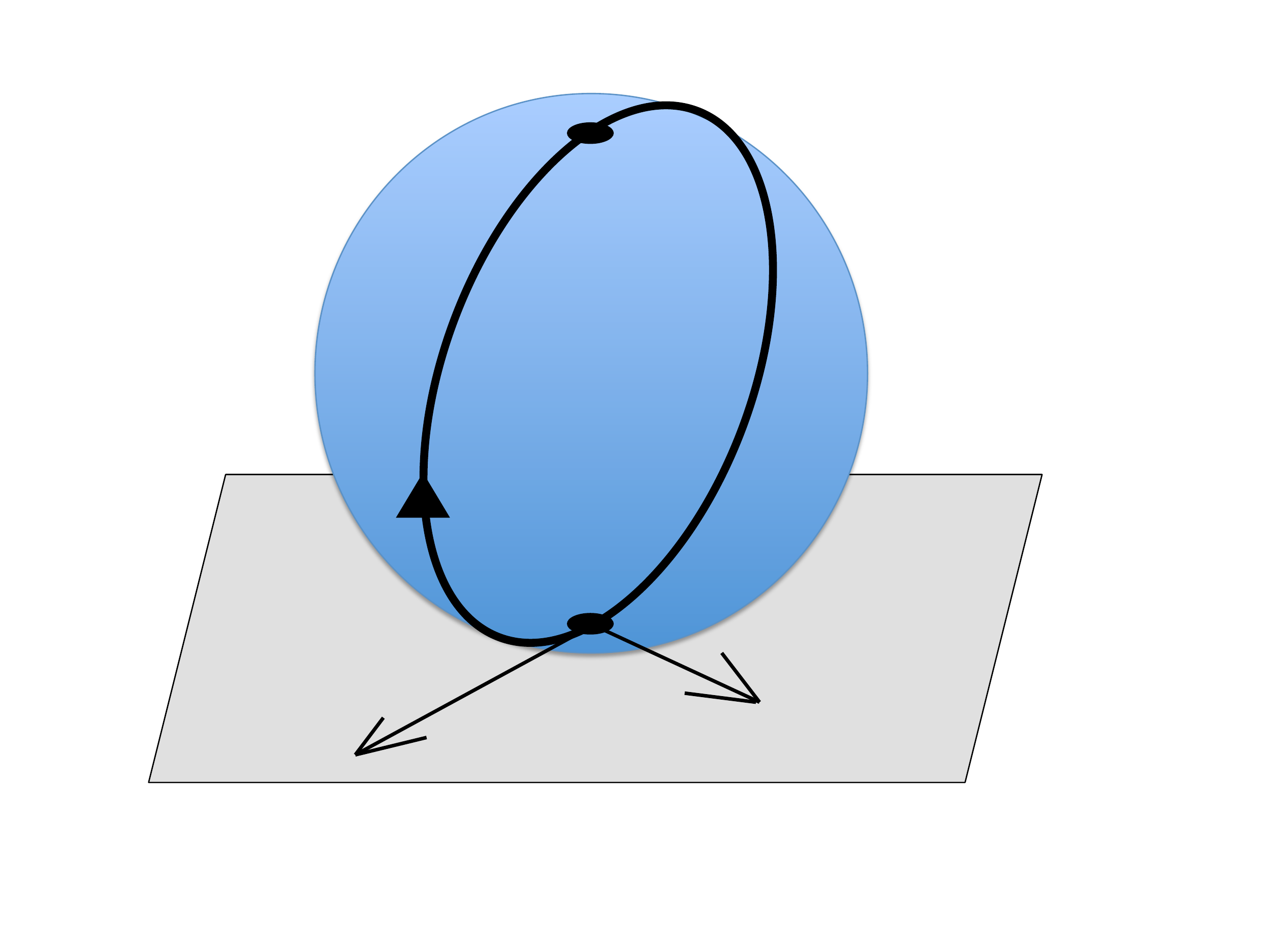}
\put(-164,100){Domain wall}
\put(-60,100){$S^3$}
\put(-150,40){$\phi_3$}
\put(-80,40){$\phi_{1,2}$}
\end{center}
\vspace{-13mm}
\caption{The topological path of the neutral pion domain wall. The sphere represents
the target space $S^3$ of the sigma model. 
On the $S^3$,  the thick line with an arrow represents the domain wall solution 
\eqref{sol}. It rounds the sphere, but topologically trivial. The plane beneath the sphere
is the parameterization space of $\phi_3$ and $\phi_{1,2}$. There is a one-to-one 
correspondence between a point on the sphere and a point on the plane.
}
\label{figs2}
\end{figure}

Finally we briefly comment on the stability of the domain walls.
Our domain wall is topologically trivial, because 
the target space of the non-linear sigma model is $SU(2) \simeq S^3$ on 
which the two vacua $\theta=0,2\pi$ are the
same point (south pole): See Fig.~\ref{figphi3} and Fig.~\ref{figs2}. 
Therefore, the domain walls can be created spontaneously,
 but on the other hand,  they can decay through the
  fluctuations along $\phi_{1,2}$ directions.
 Indeed, it was shown that 
  the domain wall is stable only for  $B_3 > 10^{19} $ [G]
 at $\beta=\tilde{\beta}=1$ by analyzing its 
  local and global stability   \cite{Son:2007ny}.
As we have shown after Eq.(\ref{b3cri}), the above condition is replaced by
$B_3 > \sqrt{\beta \gamma_0} \times  10^{19}$ [G].  This implies  
that the global stability is guaranteed for smaller $B_3$
when $\beta \rightarrow 0$.
 The analysis of the local stability is more involved especially 
   where $\beta > 0$ and $\tilde{\beta} <0$  (see Fig.\ref{fighat}): 
 In this case, the charged pion condensation
$\phi_+  =a \exp \left( -i \mu_{\pi} t +i\vec{k}\cdot \vec{x} \right)$ 
should be considered together with the 
 $\pi^0$ domain wall. We leave the analysis to our future work.\footnote{
With the condensation, the effective action for $\phi_3$ 
in Eq.\eqref{corac} has a spatial kinetic 
term with a replacement
$\beta \rightarrow (\beta - 2 \gamma_4 a^2)/(1+a^2/(4f_\pi^2))^2$.
Here $\gamma_4$ is negative ($\simeq -0.006$) at the critical $\mu_{\rm n}$ giving $\beta=0$.
So, the charged pion condensation $a\neq 0$ pushes the nearly-vanishing 
$\beta$ back to a positive nonzero,
which increases the domain wall tension. 
Hence
the charged pion tends to vanish inside the wall, and
the wall is expected to be stable against the charged pion fluctuations.}

\begin{figure}[t]
\begin{center}
\includegraphics[width=0.4\textwidth]{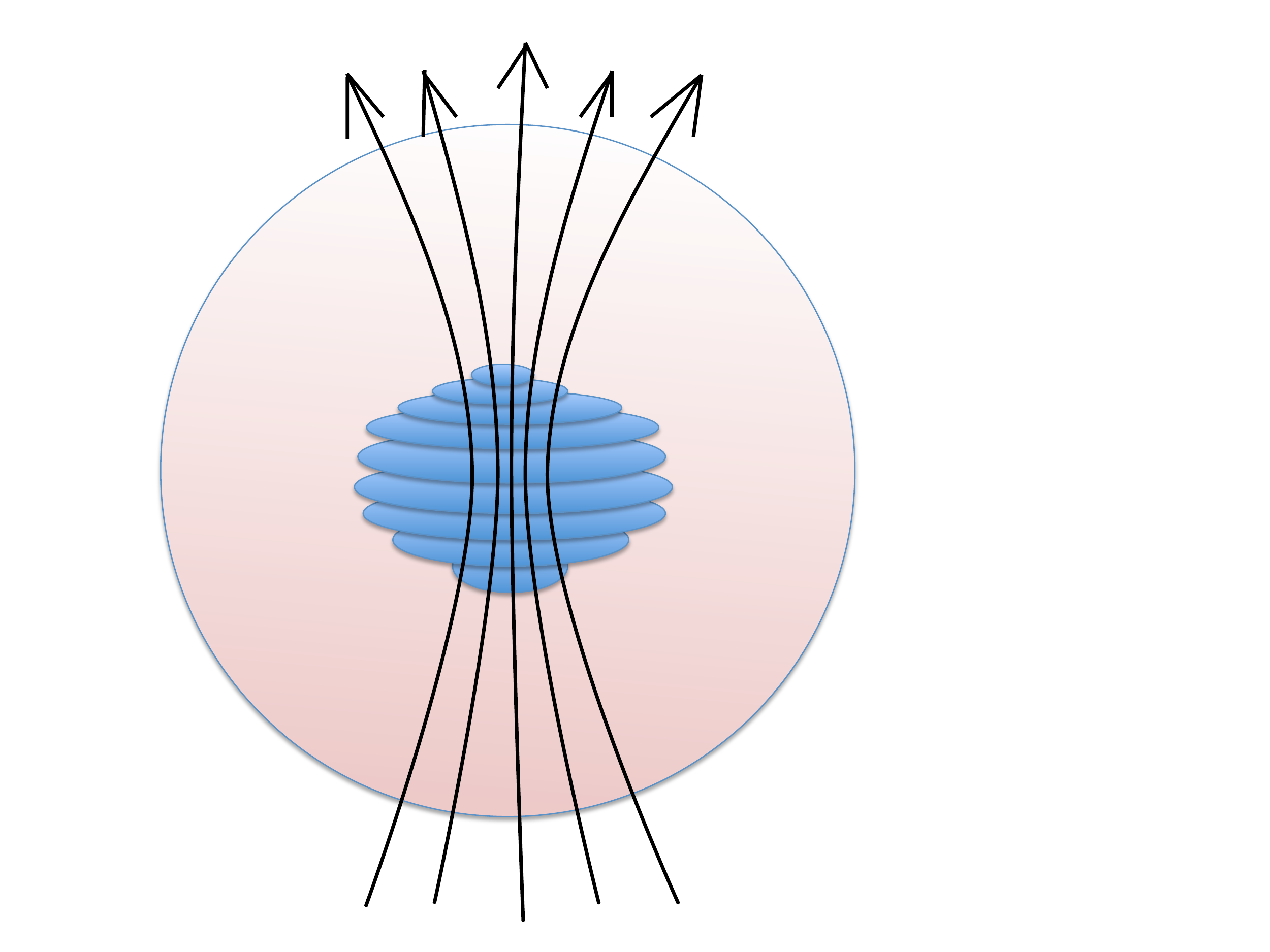}
\put(-85,130){$\vec{B}$}
\put(-90,90){Layers of neutral} 
\put(-90,80){pion domain walls}
\end{center}
\vspace{-7mm}
\caption{A schematic figure of the neutron star with domain wall layers at the core.
Scales should not be taken seriously.}
\label{figstar}
\end{figure}

\vspace{2mm}
{\noindent \bf Implication to magnetars.} ---
A schematic picture of the core region of the neutron star is shown in Fig.~\ref{figstar}.
Only at the core region, because of the high density (or rather to say the large value of
the neutron chemical potential), the domain walls are present. At the boundary of the domain
walls, neutrons drip from the wall boundary so that the total baryon number is conserved
\cite{Gorsky:2010dr}. 
As the domain wall layers, the ferromagnetic region, are present only at the core
of the neutron star, at the surface of the neutron star the magnetic field does not 
reach the value in Eq.\eqref{b3cri}, but it would be enough strong to explain the magnetars.

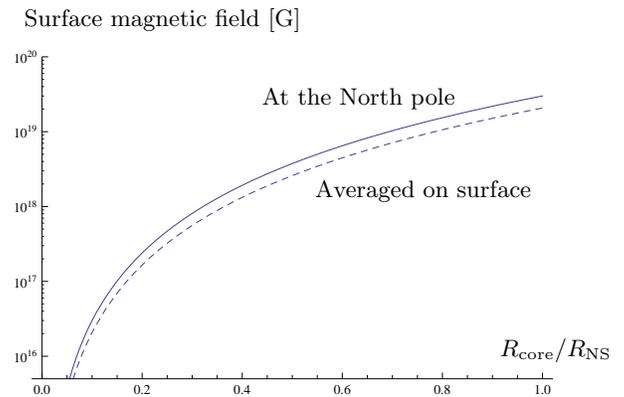
\begin{figure}[t]
\begin{center}
\includegraphics[width=0.4\textwidth]{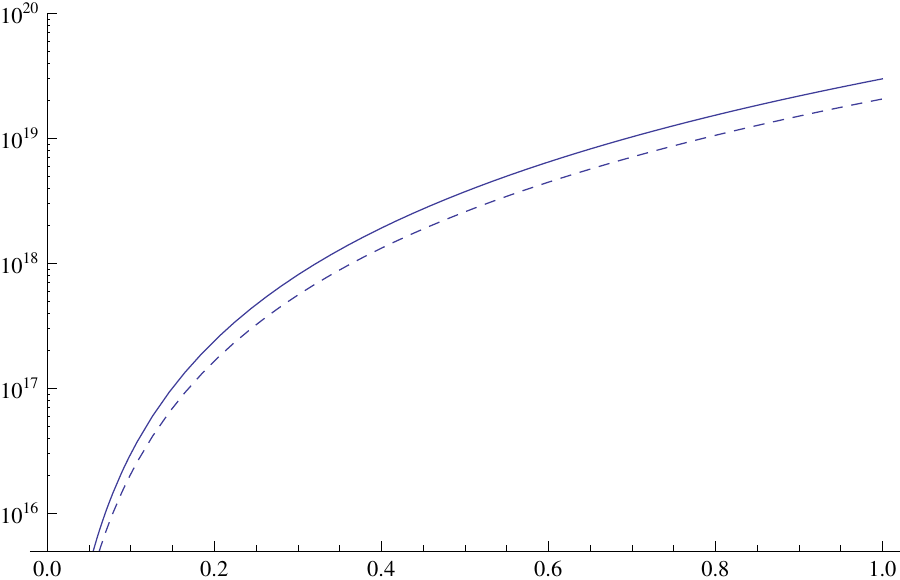}
\put(-200,140){Surface magnetic field [G]} 
\put(-110,110){At the North pole}
\put(-90,75){Averaged on surface}
\put(-20,15){$R_{\rm core}/R_{\rm NS}$}
\end{center}
\vspace{-4mm}
\caption{The magnetic field at the neutron star surface. $R_{\rm core}$ is the radius of the domain 
wall layer core (assumed to be spherical), and $R_{\rm surfce}$ is the radius of the neutron star.
At the core, the critical magnetic field \eqref{b3cri} is 
assumed. For the $1.41 M_{\odot}$ neutron star with the standard APR equation
 of state \cite{Akmal:1998cf},
 $R_{\rm core}/R_{\rm NS} = 0.1$ corresponds to the critical density of the 
 pionic wall formation $ 3.2 \rho_0$.}
  \label{figmag}
\end{figure}

The magnetic field at the surface of the neutron star is smaller than that of
the domain wall core, as $B_{\rm surface} = B_{\rm core} (R_{\rm core}/R_{\rm NS})^3$ at the North pole.
$R_{\rm core}$ is the radius of the domain wall core (assumed to be spherical and to have
a homogeneous ferromagnetism inside), and $R_{\rm NS}$ is
the radius of the neutron star. The average of the magnetic field magnitude on the neutron star surface
is $B_{\rm average} = 0.69 B_{\rm surface}$.
The core-radius dependence of the surface magnetic field is shown in Fig.~\ref{figmag}.
If the core is sufficiently small such as $R_{\rm core}/R_{\rm NS}\sim 1/10$, the surface magnetic field
may reduce to $B_{\rm surface}\sim {\cal O}(10^{16})$ [G].

The mechanism suggests that there are two kinds of neutron stars: one which reaches
the critical density and has the domain wall layer structure, and the other which does not
have it. The former would have a strong magnetic field but the latter would not have it.
It is interesting that the recent data \cite{EnotoProc} 
of the magnitude of the magnetic field on the surface 
of the neutron stars show two categories, magnetars and the others.

It is important to construct more realistic models with nuclear forces, as our model
uses free neutrons. For example, the neutron superfluidity can co-exist with the domain wall, since at higher densities the spin-aligned neutron pairing ${^3}\!P_2$ is known to be favored.
Furthermore, structure of the solitonic core of neutron stars 
would influence the equation of state, and may be sensitive to the mass-radius 
map of the neutron stars. 
If the orientation of the solitonic core is different from the rotation axis of the neutron star, it would be a source of
gravitational waves. All details need to be explored,
to match the observations of the neutron stars. We hope that our mechanism may survive
various corrections, and explain observations.

\vspace*{2mm}
{\noindent \bf Acknowledgment.} ---
The authors would like to thank T.~Enoto, D.-K.~Hong, K.~Iida, M.~Nitta, 
Y.~Suwa, T.~Tamagawa and T.~Tatsumi 
for valuable comments.
This work was initiated at a focus program in APCTP, Pohang, Korea,
and we appreciate their hospitality.
The authors are supported in part by the Japan 
Ministry of Education, Culture, Sports, Science and Technology.
This work was supported in part by
JSPS Grants-in-Aid for Scientific Research No.~22340052, 23740226, 23105716, 23654096, 22340069.


\end{document}